\documentclass[useAMS,usenatbib]{mn2e}
\usepackage{graphicx}
\usepackage[update,prepend]{epstopdf}
\title[Microstructure and polarization]{Radio-frequency microstructure and polarization in ion-proton pulsars}
\author[P. B. Jones]{P. B. Jones\thanks{E-mail:
p.jones1@physics.ox.ac.uk}  \\
University of Oxford, Department of Physics, Denys Wilkinson Building,\\
Keble Road, Oxford OX1 3RH, U.K.}

\begin{document}

\date{}

\pagerange{\pageref{firstpage}--\pageref{lastpage}}
\pubyear{}

\maketitle

\label{firstpage}

\begin{abstract}

It is shown that the time-variability inherent in the ion-proton polar cap leads naturally to growth of Langmuir modes on narrow bundles of magnetic flux lines and that the observed size of micropulses is consistent with the smallest such bundle that can plausibly support the growth of the modes.  The polarization of integrated pulse profiles is revisited, specifically, the $\pi/2$ position-angle jumps that are frequently observed accompanied by a zero in polarization.  The fact that radiation emitted above ion-proton polar caps is not tangential to the local flux lines but has a finite angular distribution proves to be the essence of understanding this phenomenon as is a constraint on the shape of the polar cap.  In principle, the behaviour should also be seen if the form of the polar-cap ion-proton area in a single pulse satisfies certain conditions which are explained here.
\end{abstract}


\section{Introduction}

The presence of quasi-periodic microstructure in the GHz emission of many pulsars (Hankins 1971) has been known for almost half a century.  Recent surveys (Kramer, Johnston \& van Straten 2002; Mitra, Arjunwadkar \& Rankin 2015) of single-pulse observations in normal pulsars show that micropulse widths are not much smaller than their repetition period $P_{\mu}$ within a single pulse so that they appear as peaks on a more slowly varying background intensity.  Over the limited interval of normal pulsar rotation periods, these authors presented some evidence that the repetition period is a linear function of $P$; very approximately $P_{\mu} \approx 1.0\times 10^{-3}P$.  However, the status of this relation has changed very recently with the detection of micropulses in two millisecond pulsars (MSP) by De, Gupta \& Sharma (2016), so extending the interval of $P$ by almost two orders of magnitude.  As De et al point out, a linear relation is strong evidence for emission sources that are moving out from the polar cap on distinct narrow bundles of magnetic flux lines successively intersected by the line of sight within the pulse profile.  There appears to be no obvious way in which the alternative source model, that micropulses represent the temporal modulation of an outward flowing plasma, could support a linear relation between $P$ and $P_{\mu}$.

More generally, there is now strong evidence that the source of the almost universal GHz radio spectra with large negative spectral indices is an ion-proton plasma (Jones  2016a,b) present in neutron stars with spin ${\bf \Omega}$ and polar-cap magnetic flux density ${\bf B}_{s}$ such that ${\bf \Omega}\cdot{\bf B}_{s} <0$ (positive polar-cap corotational charge density).

The way in which the observed circular polarization is formed in such pulsars has been described in a previous paper (Jones 2016a) but the nature of micropulse sources was not considered.  Section 2 of this brief paper describes the structure of the time-dependent ion-proton polar cap in the necessary detail.  Section 3 describes micropulse formation and compares the polarization of single and integrated pulses.
It also gives what we believe is the correct interpretation of the frequently observed $\pi/2$ position-angle discontinuities.  The conclusions are summarized in Section 4.  In this paper, the subscripts parallel and perpendicular refer to the local magnetic flux density.

\section{The ion-proton plasma}

The reverse flow of electrons at the polar cap produces electromagnetic showers and, necessarily, protons through the decay of the nuclear giant dipole state.  Maximum proton production is at a depth of about $10$ radiation lengths below the top of the neutron-star atmosphere to which region they diffuse in a time whose distribution peaks at about $1$ second. This is an unavoidable consequence of well-understood shower properties and defines the plasma composition in the ${\bf \Omega}\cdot{\bf B}_{s} < 0$ case.  Density-functional calculations of the ion cohesive energies (Jones 1985; Medin \& Lai 2006) have shown that an ion atmosphere in approximate local thermodynamic equilibrium must be presumed at the neutron-star surface.  Owing to their different charge-to-mass ratio the protons pass through and, if their flux exceeds the Goldreich-Julian value, form a layer in the atmosphere above the ions. The proton formation rate is approximately $0.2 - 0.4$ GeV$^{-1}$ electron energy, depending on $B_{s}$ and ion atomic number.

The source of reverse electrons is photo-ionization of accelerated ions by blackbody radiation from about a steradian of neutron-star surface centred on the polar cap.  This also partially screens the acceleration field ${\bf E}_{\parallel}$ in a manner completely analogous with electron-positron pair creation to the extent that ions, though relativistic, typically have modest Lorentz factors ($\gamma \leq 50$).  But the important point is that the process of reverse-electrons creating  free protons is local to any point ${\bf u}$ within the open-magnetosphere polar cap.  Transverse diffusion of protons is negligible in times no more than one or two orders
of magnitude longer than the rotation period.

We can consider the evolution of the polar cap starting at an instant at which the ion atmosphere is entirely covered by protons. A computational model (Jones 2013) has demonstrated this state and the short nulls it represents.  Protons are accelerated outwards at the Goldreich-Julian flux until at some point on the polar cap the proton layer becomes too thin to supply it.  The plasma accelerated from that point changes to ions and protons leading to a reverse-electron flux and more protons which as their flux increases eventually cut off the ion component at that point.  The condition of the complete polar cap is essentially disordered.  The ratio of the ion-proton to  proton-only areas depends on many factors, $B_{s}$ and $P$, but also the surface temperature $T^{\infty}_{s}$ and surface ion atomic number $Z_{s}$, which are not well known.

The existence of ion-proton fluxes with $\delta$-function velocity distributions at all altitudes leads to growth of a longitudinal or quasi-longitudinal Langmuir mode as described by Asseo, Pelletier \& Sol (1990), nonlinearity and turbulence.  Two components with differing charge-to-mass ratios and hence different velocities are essential to the growth of such a mode.  Thus the development of plasma turbulence is restricted to bundles of flux lines with ion and proton components.  Regions of polar cap emitting only protons produce no unstable mode and no coherent radio emission.  We refer to Jones (2016b) and papers cited therein for more complete details of this work.

\section{The polar cap and micropulses}

The relation between source electron-positron Lorentz factors and micropulse size has been studied empirically by Luo (2004) in the context of the pair-creation polar cap model.  He represented the angular width of the radiation distribution by $\chi = \gamma^{-1}$ and found micropulse widths consistent with $\gamma \sim 10^{2}$.  The ion-proton polar cap leads to a specific relation between $\chi$ and $\gamma$ in equation (2) below which gives smaller values of $\gamma$ for the ions and protons.

The shape of the active polar cap supporting plasma turbulence is determined by the magnetosphere in the region of the light cylinder radius $R_{LC}$ and the angle $\psi$ between magnetic and rotation axes.  It is unknown in detail, but is usually represented by a circular or semi-circular approximation.  Here, we shall find these specific approximations inadequate and therefore initially assume some general shape, as shown in Fig. 1, with dipole flux lines defining its circumference, eccentrically positioned with respect to the magnetic axis $M$.
Our assumption is that plasma turbulence couples with the radiation field and that we can define, for practical purposes, a surface of last absorption at radius $\eta_{e}$ beyond which radiation propagates as the O- and E-modes of a birefringent open magnetosphere.  The angles subtended by the tangents to the flux lines at $\eta_{e}$ lie inside the solid line boundary of Fig. 1 whose extremities are at an angle of roughly $3u_{0}\eta^{1/2}_{e}/2R$ with respect to the magnetic axis.  The polar coordinate radius $\eta$ is in units of the neutron-star radius $R$ and $u_{0}$ is the approximate radius of the polar cap on the neutron-star surface.  The angles of photons emitted at $\eta_{e}$ are preserved at $\eta > \eta_{e}$ but the flux-line tangents are larger there by $(\eta/\eta_{e})^{1/2}$ as indicated by the broken-line boundary in Fig. 1.  (Aberrations are small, except possibly for MSP, and have been neglected.)

\subsection{Micropulses}

In the rest frame of an element of turbulent plasma moving with Lorentz factor $\gamma$, the photon energy density is assumed to be isotropic and at frequency $\nu_{c}$ is,
\begin{eqnarray}
\nu_{c} N_{c} & = & A(\nu_{c}/\nu_{c0})^{-\alpha} \hspace{1cm} \nu_{c} > \nu_{c0} \nonumber \\
  & = &  0    \hspace{25mm} \nu_{c} < \nu_{c0}
\end{eqnarray}
the cut-off being related loosely to the terminal unstable mode frequency.  Micropulses are defined by the Lorentz transformation to the observer frame,
\begin{eqnarray}
\nu N(\nu,\chi) = A\gamma^{2 + \alpha}\left(\frac{\nu}{\nu_{c0}}\right)^{-\alpha}
\left(\frac{2}{1 + \gamma^{2}\tan^{2} \chi}\right)^{2 + \alpha},
\end{eqnarray}
in which $\chi$ is the angle between the radiation and the direction of motion of the emitting element of plasma, and subject to there being a minimum value of $\chi$ given by $1 + \gamma^{2}\tan^{2}\chi_{m} = 2\gamma\nu_{c0}/\nu$.  (We have to note that the final exponent in equation (2) has previously been given incorrectly as $3 + \alpha$ in Jones 2016a.) The radiation from a narrow bundle of flux lines is shown schematically on the surface $\eta$ in Fig. 1.  Radiation sources at any instant can be either substantial areas of polar cap or small growing areas as described in Section 2.  Micropulses are the latter.

\begin{figure}
\includegraphics[trim = 5mm 30mm 5mm 100mm,clip,width=84mm]{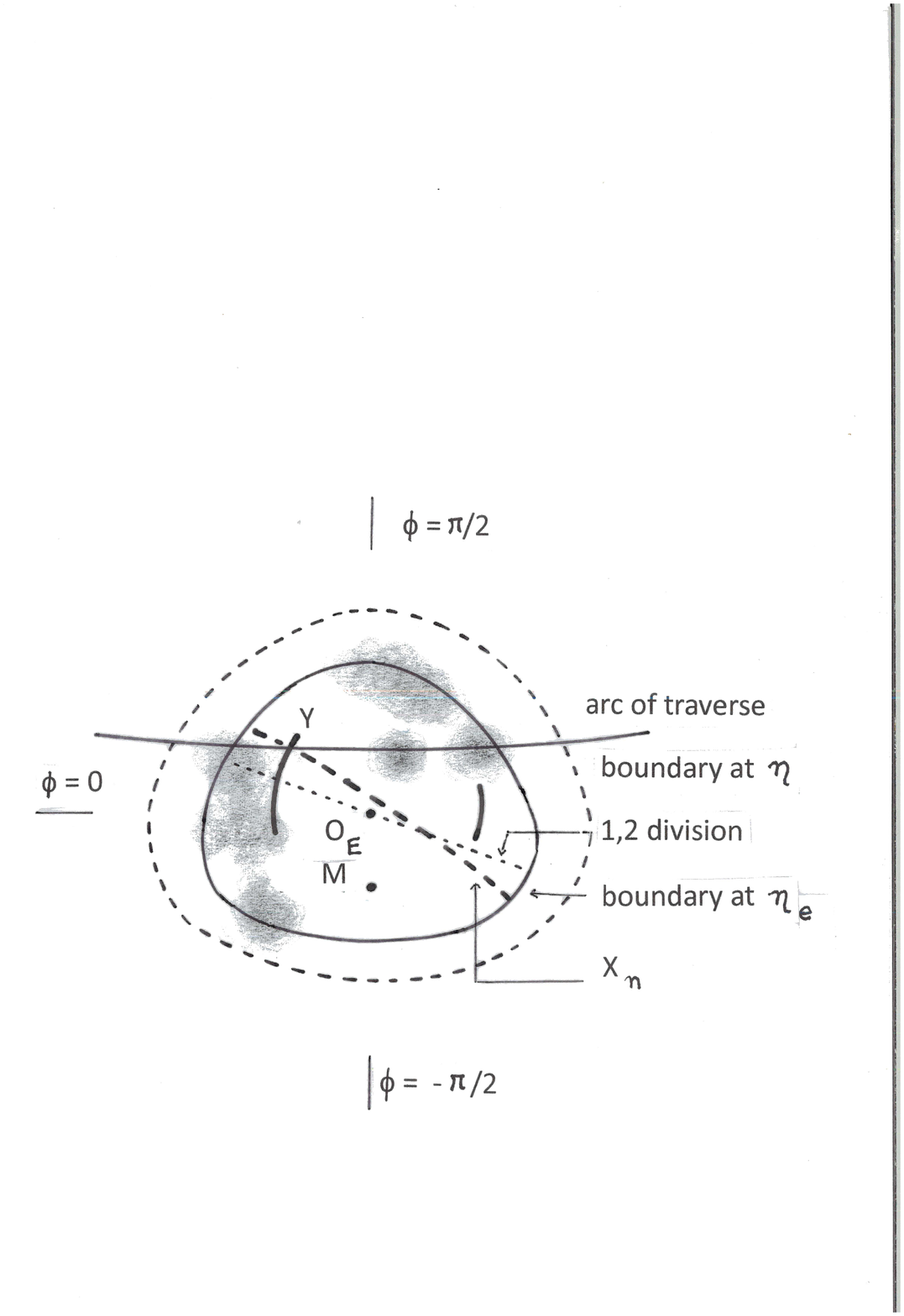}

\caption{The solid-line boundary encloses a section of the open magnetosphere at the surface of last absorption, altitude $\eta_{e}$.  Points within it give the polar and azimuthal angles of tangents to the flux lines at $\eta_{e}$.  The boundary is schematic and is eccentrically placed with respect to the magnetic axis $M$.  The outer broken-line boundary is that for tangents at $\eta > \eta_{e}$. The position angle of source elements is that of the local electrostatic field ${\bf E}_{\perp}$ above the polar cap.
The point $O_{E}$ of zero $E_{\perp}$ may be displaced from the magnetic axis $M$ and is the origin to which the polar and azimuthal angles are referred. Radiation emitted on a flux line at $\eta \approx \eta_{e}$ has angular distribution given by equation (2) with respect to that line and is shown schematically as it would be for a single pulse by grey shading. In the case of integrated profiles, an example of the notional curves $X_{n}$ used in the construction of the loci of zeros in $p$ is shown schematically, as are the loci themselves, represented by heavy-line segments.  Their intersection with the arc of traverse of the sight line at $Y$ gives the longitude of the observed $\pi/2$ discontinuity.}

\end{figure}

We adopt $\alpha = 2$ in equation (2) as indicated by the observed spectral indices.  Then $\chi \approx \gamma^{-1}(f^{-1/4} - 1)^{1/2}$ is the angle at which the micropulse intensity is a fraction $f$ of the maximum.  Only the peaks of micropulses are usually observable, the lower parts merging into a smoothly varying background so that for $f = 1/2$, the angle is $\chi = 0.43\gamma^{-1}$.
Micropulse widths are in most cases not well-determined.  Therefore we compare $\chi$ with the angle $10^{-3}(2\pi\sin\psi)$ defined by $P_{\mu}$, where $\psi$ is the angle between the magnetic and rotational axes.   With the elimination of $\chi$ we find $\gamma \approx 69\csc\psi$. The values of $\gamma$ indicated by this comparison are at the high end of the interval expected and the $\csc\psi$-dependence increases the discrepancy. But the distribution of the alignment angle $\psi$ and the process of plasma turbulence formation and decay are not well understood so that such a discrepancy may not be a serious problem.

One obvious question is how narrow can a micropulse be?  The descriptions of Langmuir mode growth have been strictly one-dimensional, but a column of ions and protons moving on a narrow bundle of flux lines embedded in outward moving protons is certainly not one-dimensional.  We do not know how its width affects the growth rate of the mode, but there must be an effective minimum.  The best known micropulse widths are those of the Vela pulsar (Kramer et al 2002; Fig. 3).  There is a distinct correlation of width with mean peak flux density, and a cut-off at $\sim 50$ $\mu$s, which in principle is entirely consistent with our statement.  Expressed as an angle, the cut-off is at $3\times 10^{-3}\sin\psi$.  At a height of $\eta_{e} = 5$, this is a transverse length of $2\times 10^{4}\sin\psi $ cm and at $\nu_{c0} = 5$ MHz, chosen to give a low-frequency spectral turn-over at $\nu < 100$ MHz, this would be equivalent to two wavelengths, a plausible minimum width that would support a mode growth-rate approaching the true one-dimensional value. We are obliged to presume that finer detail in single pulses must be temporal in origin and a consequence of the stochastic nature of the turbulence.  This would not be surprising in view of the very small volumes that are concerned.  The ion-proton polar cap predicts that micropulse widths should be frequency-independent, as found by Kramer et al, but gives no immediate explanation for quasi-periodicity.

But the concept of quasi-periodicity appears somewhat stretched here: the distribution of auto-correlation function first maxima given in Fig. 2 of Kramer et al does not at first sight suggest it.  Certainly the authors quote large errors in $P_{\mu}$.  Also the micropulse widths are a large fraction of $P_{\mu}$.  For these reasons, the failure to predict quasi-periodicity is not disturbing.

\subsection{Integrated and single-pulse polarization}

There are many published studies of circular and linear polarization in integrated pulse profiles.  Two features are particularly significant.

(i)  The polarization position-angle may undergo a discontinuity of $\pi/2$ within an extremely small interval of longitude, for example, one bin in the high-resolution observations of Karastergiou \& Johnston (2006), and in coincidence with a zero in polarization $p$.

(ii) Circular polarization varies smoothly within the integrated profile and changes sense both in coincidence with the position-angle discontinuities and elsewhere, particularly near the centre of the profile.

Single-pulse polarization differs in that it is more difficult to observe in normal pulsars (Kramer et al 2002, Mitra et al 2015) or in MSP (De et al 2016).  The fine detail present is presumably temporal and stochastic.

Earlier work on these problems was empirical in nature and we refer to Melrose et al (2006) who also review the relevant literature.  An empirical view of polarization phenomena led to the model of two approximately orthogonal polarization modes (OPM) formed from an array of subsources.  Their intensities are partially correlated and they follow different ray paths (a secondary electron-positron plasma with a substantial difference between the refractive indices of its normal modes is implicit).  Gaussian distributions are postulated for the mean and variance of the relative intensity of the modes and for the means and variances of their position on the Poincar\'{e} sphere.  Distinct physical modes are assumed but with no explanation of their formation.

However, the division between two sectors made here in this Section is notional: it is a technique to establish the existence of $p = 0$ loci on the polar cap.  There is no physical distinction between any of its parts.  The existence and extent of $p = 0$ loci is dependent only on the function $A({\bf u},t)$ which represents those areas of the polar cap that are emitting ions and protons at any instant.

This is directly relevant to any distinction between integrated profiles and single pulses.  In principle, there is none.  If the position-angle jumps are less frequently observed in single pulses, it is merely because the form of $A({\bf u},t)$, for example,  as shown here schematically in Fig. 1 with $A$ negligibly small over much of the polar cap, may not lead to $p = 0$ loci of an extent that makes intersection with the arc of traverse probable.

These phenomena have a clear physical basis in the ion-proton polar cap.  The formation of polarization can be divided into two successive stages.  Firstly, the radiation produced when a turbulent charge-density fluctuation interacts with the electromagnetic field  is usually linearly polarized.  We shall assume its position angle at any point is approximately that of ${\bf E}_{\perp}$, the electrostatic field above the polar cap, whose origin $O_{E}$ may not be on the magnetic axis.  Thus away from the centre, it is roughly perpendicular to the boundary.  This assumed position angle is the basis of the description of the polar cap given here but it is not inconsistent with the overall longitude-dependence of polarization position-angle first described by Radhakrishnan \& Cooke (1969) on which the $\pi/2$ jumps are superimposed in some pulsars. A further minor point is that it is also approximately the position angle ${\bf k}\times({\bf k}\times{\bf B})$ of the electric field in the quasi-longitudinal Langmuir mode described by Asseo et al (1990)for wavevector ${\bf k}$ in the plane of flux-line curvature. This mode, as opposed to the longitudinal mode, does couple directly with the radiation field. But we nevertheless assume that the development of turbulence is predominantly the process through which this coupling occurs.

We require the Stokes parameters at the surface of last absorption $\eta_{e}$ for radiation of unit wave-vector $\hat{\bf k}(t)$ as the line of sight moves  on the arc of traverse, and these are obtained by integration over all flux lines passing through the surface using the distribution of $\chi$ given by equation (2). The distribution of ion-proton areas on the polar cap is represented by the function $A({\bf u},t)$.  Thus in Fig. 1, representing a single pulse, the shaded areas near the arc of traverse are those contributing predominantly to the single pulse. In the case of integrated pulse sequences, the function $A$ will be smoothly varying on $\eta_{e}$ though not necessarily uniform owing to the variation of $\gamma$ with position on the polar cap.

Having found the Stokes parameters for wave-vector $\hat{\bf k}$ at $\eta_{e}$, the linear polarization can be resolved into a superposition of the O- and E-modes of the outward flowing plasma at $\eta > \eta_{e}$.  The important quantity is the difference between the O- and E-mode refractive indices integrated from $\eta_{e}$ to infinity.  The birefringence, which is a function of the ion-proton masses and Lorentz factors, leads to a circularly polarized component which is predicted to vary slowly with longitude.  This is what is observed and is strong evidence that the birefringence is small and can be sourced only by baryonic-mass particles, not electrons or positrons.  We refer to Jones (2016a) for a more complete account.

The $\pi/2$ discontinuities in position angle are a common feature of integrated profiles.  It is natural to refer to the polarization formed by the addition of the Stokes parameters for two incoherent sources of intensity $I_{1,2}$, polarizations $p_{1,2}$ and position angles $\phi^{p}_{1,2}$.  The resultant polarization is $p$,
\begin{eqnarray}
p^{2} = \frac{|I_{1}p_{1} - I_{2}p_{2}|^{2} + 2I_{1}I_{2}p_{1}p_{2}
 (1 + \cos(2\phi^{p}_{1} - 2\phi^{p}_{2}))}{(I_{1} + I_{2})^{2}},
\end{eqnarray}
and the necessary conditions for $p = 0$ are $I_{1}p_{1} = I_{2}p_{2}$ and $\phi^{p}_{1} - \phi^{p}_{2} = \pi/2$.  This well-known result is useful in relation to the observed position-angle discontinuities.

Summed over many single pulses, as in the integrated profile, the function $A({\bf u})$ is time-independent and smoothly varying over the polar cap and the surface at $\eta_{e}$.  We can arbitrarily divide the polar cap on the neutron-star surface into two notional sectors (1,2) of roughly equal area each being a plasma source whose Stokes parameters found by integration over the sector give overlapping distributions of radiation on the surface at $\eta_{e}$.
Then it is always possible to define a curve $X_{n}$ on which $I_{1}p_{1} = I_{2}p_{2}$ whose shape is dependent on $A({\bf u})$ and $\gamma$ and is generally different from the (1,2) line of division.  From the Stokes parameters for the two sectors, it is evident that $\phi^{p}_{1} - \phi^{p}_{2} \approx 0$ at the ends of $X_{n}$.  Provided the position angles on $X_{n}$ satisfy $\pi/2 < \phi^{p}_{1} - \phi^{p}_{2} < \pi$ in the limit as $O_{E}$ is approached, the $p = 0$ condition $\phi^{p}_{1} - \phi^{p}_{2} = \pi/2$ must be satisfied at two points on $X_{n}$.  In terms of the Stokes parameters this is, $Q_{1} + Q_{2} = U_{1} + U_{2} = Q + U = 0$.  Because the (1,2) division is arbitrary, it can be repeated indefinitely to give an infinite set of $X_{n}$, so mapping out segments of curves on the surface at $\eta_{e}$ which are the loci of the zeros in $p$.  These arrangements are shown schematically in Fig. 1.

The factors which determine the limit behaviour of the position angle difference are demonstrated by direct calculation of the Stokes parameter $Q$ on the symmetry axis of a model polar cap consisting of two semi-circles of radius $\theta_{0} = 3u_{0}\eta^{1/2}_{e}/2R$, separated by a rectangle of area
 $2\theta_{0} \times 2\zeta$.  The procedure is precisely as described in this Section prior to equation (3). Here $u_{0}$ is the polar-cap radius on the neutron-star surface and  the simple shape defined here approximates an ellipse with semi-major axis $\theta_{0}+\zeta$ and semi-minor axis $\theta_{0}$. Both $A({\bf u})$ and $\gamma$ are independent of position.  In this model ,the local electric field and polarization vectors are completely defined by the assumption that they are rectilinear and perpendicular to the boundary of the open region.  Then $U_{1} + U_{2} = 0$ on the semi-major axis and, for significantly wide intervals of $\gamma$ and $\zeta$, computation shows that $Q_{1} + Q_{2} = 0$ at two points on that axis.  These intervals are, very approximately, $1/2\theta_{0} < \gamma < 2/\theta_{0}$ and $\theta_{0}/8 < \zeta < \theta_{0}/4$.  A circular polar cap with $\zeta = 0$ does not have zeros in $Q$: the essential polar-cap shape has, roughly, some degree of ellipticity (from which our model does not differ much) with the semi-major axis nearly parallel with the arc of traverse. For a typical $\theta_{0} = 0.04$ at altitude $\eta_{e} = 5$, $\gamma \approx 1/\theta_{0} = 25$.

Intersection of one or more of these loci by the arc of traverse (the point $Y$ in Fig. 1) produces corresponding zeros in the longitudinal distribution of $p$ and their presence is also a sufficient condition for the coincident $\pi/2$ position-angle jumps, provided the arc of traverse is not locally parallel with $X_{n}$ at $Y$.  It is obvious why two zeros are not infrequently seen, although the scheme in Fig. 1 shows a single $\pi/2$ jump at point $Y$. The existence and extent of these loci on a real polar cap depends on $A({\bf u})$  and $\gamma$ which are both position-dependent.    With the benefit of hindsight, it can be seen that the form of polar cap chosen in previous work (Jones 2016a Figs. 2 and 3) was unfortunate and does not readily adapt to the arguments given here following equation (3).

Discontinuities of $\pi/2$ are not ubiquitous but are sufficiently frequent, as are double discontinuities, to require a simple explanation as given here, particularly as they occur within a single high-resolution bin of longitude.  The essential feature which makes this possible is the non-locality caused by the finite values of $\chi$.  The position angles $\phi^{p}_{1,2}$ at a point on $X_{n}$ are given by summation over the Stokes parameters of many elements of source over an area determined, through equation (2), by the Lorentz factor $\gamma$.  If it were assumed that at emission, radiation is always exactly tangential to the local flux lines, the position angle at a point on the arc of traverse would be simply that of the source element associated with that point.  As a function of longitude the observed position angle would then follow the classical form described by Radhakrishnan \& Cooke (1969).  Even with finite $\chi$, this is observed in some pulsars, such as Vela, where the positional angle has the classical rotation but with no zero in linear polarization.  In these cases, the arc of traverse evidently crosses the polar cap near the boundary and so does not intersect any of the $p = 0$ loci.

\section{Conclusions}

Many authors have observed that the characteristics of GHz radio emission, including polarization and the large negative spectral indices, are broadly similar over wide intervals of surface magnetic flux density, from $10^{8}$ G in the MSP to more than $10^{13}$ G.  This should be in itself sufficient to prompt questions about theories of the emission process that are based on electron-positron pair production.  The very recent addition of microstructure to the list by De et al (2016) reinforces this.  The ion-proton polar cap is concomitant with the plasma composition that basic nuclear interactions must produce in ${\bf \Omega}\cdot{\bf B}_{s} < 0$ neutron stars.  It does not involve electron-positron pair creation and can function at all known polar-cap fields (see Jones 2016b and papers cited therein).  It does give an understanding of the polarization that is observed in integrated profiles, in particular, the circular polarization whose characteristics show that the limited degree of birefringence of the magnetosphere between source and observer is strong evidence for a plasma of baryonic-mass particles (Jones 2016a).

The present paper has attempted to give an understanding of microstructure in terms of the ion-proton polar cap.  It also revisits the $\pi/2$ discontinuity in position angle which is frequently observed in the integrated profile and offers a general explanation that is valid for plasma Lorentz factors of an order of magnitude that both allows growth of the Langmuir mode and is not too inconsistent with microstructure widths.

The reservations about this work are firstly, that the progression from ion-proton Langmuir-mode growth to non-linearity and turbulence and its decay requires much more study with attention, in particular, to the three-dimensional nature of the system.  Secondly, we have made a specific assumption about the linear polarization of the radiation in turbulence decay.  Apart from these, the ion-proton polar cap is based on factors each having a fairly transparent physical basis and we believe that it does offer at least a qualitative understanding of the pulsar emission process.

\section*{Acknowledgments}

I thank Dr Aris Karastergiou for some very helpful comments on polarization position-angle phenomena and for introducing me to the work of Melrose et al.

.

\bsp

\label{lastpage}


\begin{thebibliography}{99}
\bibitem{b1}Asseo E., Pelletier G., Sol H., 1990, MNRAS, 247, 529
\bibitem{b2}De K., Gupta Y., Sharma P., 2016, ApJ, 833, L10
\bibitem{b3}Hankins T. H., 1971, ApJ, 169, 487
\bibitem{b4}Jones P. B., 1985, Phys. Rev. Lett., 55, 1338
\bibitem{b5}Jones P. B., 2013, MNRAS, 431, 2756
\bibitem{b6}Jones P. B., 2016a, MNRAS, 455, 3814
\bibitem{b7}Jones P. B., 2016b, MNRAS, 459, 3307
\bibitem{b8}Karastergiou A., Johnston S., 2006, MNRAS, 365, 353
\bibitem{b9}Kramer M., Johnston S., van Straten W., 2002, MNRAS, 334, 523
\bibitem{b10}Luo Q., 2004, MNRAS, 352, 1208
\bibitem{b11}Medin Z., Lai D., 2006, Phys. Rev. A, 74, 062508
\bibitem{b12}Melrose D., Miller A., Karastergiou A., Luo Q., 2006, MNRAS, 365, 638
\bibitem{b13}Mitra D., Arjunwadkar M., Rankin J. M., 2015, ApJ, 806:236
\bibitem{b14}Radhakrishnan V., Cooke D. J., 1969, Astrophys. Lett., 3, 225

\end{thebibliography}
\end{document}